\documentclass[
12pt,
nofootinbib,
amsmath,amssymb,
aps,
prd,
floatfix,
]{revtex4-1}

\usepackage[utf8x]{inputenc}
\usepackage{setspace}
\usepackage{lmodern}
\usepackage[T1]{fontenc}
\usepackage{graphicx}
\usepackage{dcolumn}
\usepackage{bm}
\usepackage{verbatim}
\usepackage{url}
\usepackage{hyperref}
\usepackage{natbib}
\usepackage{etoolbox}
\usepackage{tikz}
\usepackage{longtable}
\usepackage{multirow}
\usepackage{siunitx}
\usepackage[export]{adjustbox}
\usepackage{array}
\newrobustcmd*{\tikzsquare}[1]{\tikz{\filldraw[draw=#1,fill=#1] (0,0) rectangle (0.6em,0.6em);}}%
\newrobustcmd*{\tikztriangle}[1]{\tikz{\filldraw[draw=#1,fill=#1] (0,0) -- (0.2cm,0) -- (0.1cm,-0.2cm);}}

\begin{document}
\singlespacing
\title{Large-scale simulations of antihelium production in cosmic-ray interactions}

\author{Anirvan Shukla}
\email{anirvan@hawaii.edu}
\author{Amaresh Datta}
\author{Philip von Doetinchem}
\author{Diego-Mauricio Gomez-Coral}
\affiliation{Department of Physics and Astronomy, University of Hawai'i at M\={a}noa, 2505 Correa Road, Honolulu, Hawaii 96822, USA}

\author{Carina Kanitz}
\affiliation{Erlangen Center for Astroparticle Physics, Friedrich-Alexander-Universität Erlangen-Nürnberg, Erwin-Rommel-Str. 1, 91058 Erlangen, Germany}

\date{\today}

\begin{abstract}
The possibility of antihelium production in interaction of cosmic rays with the interstellar gas is studied using large-scale Monte Carlo simulations. For this purpose, an energy-dependent coalescence mechanism developed previously is extended to estimate the production of light antinuclei (${}^3\overline{\text{He}}$ and ${}^4\overline{\text{He}}$). The uncertainty in the coalescence parameter and its effect on the expected antiparticle flux is also investigated. The simulated background antihelium fluxes are found to be lower than the fluxes predicted by simplified models using numerical scaling techniques. 
\end{abstract}

\maketitle

\section{Introduction}\label{s0}

Antideuterons and antihelium nuclei are a potential breakthrough approach for dark matter searches because dark matter induced cosmic-ray (CR) antinuclei fluxes predicted by many different models exceed the estimated astrophysical background in the energy range of GeV or sub-GeV by orders of magnitude~\cite{doetinchem2020cosmicray, Donato:1999gy,Baer:2005tw,Donato:2008yx,Duperray:2005si,Ibarra:2012cc,Ibarra2013a,Fornengo:2013osa,Dal:2014nda, Korsmeier:2017xzj,Tomassetti:2017qjk,Lin:2018avl,Li:2018dxj,Cirelli:2014qia, Carlson:2014ssa, Coogan:2017pwt}. In our matter-dominated Universe, astrophysical production of antimatter can occur only as pair production from the collision of cosmic rays with interstellar matter (ISM) particles, with protons being the largest component of both the CR and ISM (in the form of hydrogen gas). Antinuclei can be formed in collisions with energy above their respective production thresholds. This threshold for light antinuclei increases steeply with antinucleon number because every additional antinucleon requires the production of a corresponding nucleon as well. The energy thresholds for $\overline{\text{\textit{d}}}$, ${}^3\overline{\text{He}}$, and ${}^4\overline{\text{He}}$ in $p$-$p$ interactions are about 17, 31, and 49\,GeV, respectively, in the target frame or about 5.7, 7.5, and 9.7\,GeV, respectively, in the center-of-mass frame. 

Within the first few years of operation, the space-based AMS-02 experiment entered the precision era for cosmic-ray antiproton measurements~\cite{PhysRevLett.117.091103} and recently reported several antihelium candidate events~\cite{antihe, antihe2}. Naively, this leads to the assumption that antideuterons should be observable in large quantities as well. However, thus far, no strong antideuteron candidates have been reported by the AMS-02 Collaboration. These unexpected antihelium observations have spurred an interest in studying the secondary production and propagation of antihelium in our Galaxy. Most of these semianalytical studies have relied on simplified numerical scaling of antiproton production cross sections to predict the production cross sections of heavier antinuclei in typical CR-ISM interactions~\cite{PoulinVivianSalatiPhysRevD.99.023016,Korsmeier:2017xzj,PhysRevD.68.094017}.

This study tries a different approach by using an event-by-event implementation of the coalescence model~\cite{PhysRev.129.836, CSERNAI1986223, BALTZ19947, GomezCoral:2673048, Ding_2019}. In Ref.~\cite{Gomez-Coral:2018yuk}, antideuteron production cross section measurements were fitted with simulations to determine the best-fit coalescence momentum parameter $p_0$ for proton-proton collisions at different kinetic energies. The $p_0$ was found to be energy dependent. Compared to analytical models which use a constant $p_0$, this approach can lead to important differences in the final predicted particle fluxes. In this study, the new parametrization was used to further develop a multiparticle coalescence mechanism. This approach benefits from the continuous improvement of Monte Carlo (MC) particle interaction simulators; the development of an event-by-event afterburner; and, finally, the availability of high-throughput computational facilities. Utilizing massive computation power of about 5,000 years of CPU time, more than 25 trillion proton-proton collisions were simulated at different collision energies. The total number of $p$-A collisions simulated in this study is a few orders of magnitude more than what was feasible just a few years ago.

The antitriton and ${}^3\overline{\text{He}}$ yields from this simulation were validated by comparing them to available accelerator data. This is also the first MC simulation to predict ${}^4\overline{\text{He}}$ yields, which can be compared to data from future experiments. This model could be useful in describing the formation of light antinuclei in a variety of systems for a large range of energies using a single energy-dependent coalescence parameter.

\section{Coalescence Formation of Light Antinuclei}\label{s1}

\subsection{Coalescence of two antinucleons}\label{s1ss1}
The production mechanism of light antinuclei from hadronic interactions is not well understood. A number of models attempt to describe this process. One of these is the coalescence model, which has been successful in describing the light antinuclei formation so far, as the ALICE and other results have shown~\cite{Gomez-Coral:2018yuk}. In the simple coalescence model, the fusion of an antiproton and an antineutron into an antideuteron is based on the assumption that any antiproton-antineutron pair within a sphere of radius $p_0$ in momentum space will coalesce to produce an antinucleus. The coalescence momentum $p_0$ is a phenomenological quantity and has to be determined through fits to experimental data~\cite{PhysRev.129.854}. In this approach, the antideuteron spectrum is given by
\begin{equation} \label{eq:iso_coalescence}
\gamma_d\frac{\text d^3N_{\overline{d}}}{\text d p^3_{\overline{d}}}=\frac{4\pi}{3}p_0^3\left(\gamma_p\frac{\text d^3N_{\overline{p}}}{\text d p^3_{\overline{p}}}\right)\left(\gamma_n\frac{\text d^3N_{\overline{n}}}{\text d p^3_{\overline{n}}}\right),
\end{equation}
where $p_i$ and $\text dN_i/\text dp_i$ are, respectively, the momentum and the differential yield per event of particle $i$ ($\overline{d}=$ antideuteron, $\overline{p}=$ antiproton, $\overline{n}=$ antineutron). This is known as the analytical coalescence model. However, this is overly simplistic since it does not take into account effects like energy conservation, spin alignment etc., which have an important effect on deuteron and antideuteron formation. It also assumes that the production of antiprotons and antineutrons is uncorrelated~\cite{Chardonnet_1997} and expresses the momentum distribution of the coalesced particle as the product of two independent isotropic distributions. This is another simplification since correlations have an important effect on the coalescence process~\cite{alej2013determination, alej2012prospects, kadastik2009enhanced}.

\par To take into account the hadronic physics (energy and momentum, angular correlations, event topography, antiproton-antineutron production asymmetry, etc.), MC hadronic event generators are used. Typical hadronic generators~\cite{Pierog:2013ria, Ostapchenko:2004qz, Sjostrand:2007gs, Agostinelli2003250, 2006ITNS53270A, Galoyan:2012bh, Ahn:2009wx} do not have the capability to produce (anti)deuterons. Therefore, the state-of-the-art technique is to create an event-by-event coalescence model afterburner coupled to the hadronic generators. The afterburner applies the coalescence condition to $\overline{\text{\textit{p}}}\overline{\text{\textit{n}}}$ or $pn$ pairs on a per-event basis. (e.g., Refs.~\cite{Kadastik:2009ts, Ibarra2013a, Gomez-Coral:2018yuk}). For each event, the momentum difference of each antinucleon pair is calculated in their corresponding center-of-mass frame. If the momentum difference is smaller than $p_0$, a new particle is produced with a momentum equal to the sum of the constituent particle's momenta. The coalescence condition can be expressed as
\begin{equation}
\lvert \vec{k}_{\bar{p}} - \vec{k}_{\bar{n}} \rvert < 2p_0
\label{eqn:coalescenceCondition}
\end{equation}
The coalesced particle's binding energy is taken into account by calculating its total energy from its calculated momentum and the Particle Data Group~\cite{PhysRevD.98.030001} value of its rest mass. The constituent antiprotons and antineutrons are removed from the event, and the process is repeated for all remaining antinucleons, until all possible pairs are exhausted. 
The coalescence momentum $p_0$ is varied as a free parameter, and best-fit values are obtained by comparisons with the experimental data. It is important to note that the coalescence model is not a nuclear-physics model for the formation of light antinuclei from first principles. It should be seen as an empirical approach that is capable of reproducing the experimental data.

\subsection{Choice of the Monte Carlo event generator}\label{s1ss3}
To simulate the production of antideuterons and larger antinuclei, an accurate description of the production of constituent particles (antiprotons and antineutrons) is of paramount importance. Previously, in Ref.~\cite{Gomez-Coral:2018yuk}, the formation of deuterons and antideuterons was studied using multiple MC event generators in the framework of Cosmic Ray Monte Carlo package ({\tt CRMC})~\cite{crmcRef}. It was demonstrated in that study that the parametrization of $p_0$ depended on the choice of the MC event generator. Further, the {\tt EPOS-LHC} Monte Carlo event generator~\cite{Pierog:2013ria} was shown to be consistent with $\overline{\text{\textit{p}}}$ production data in a wide range of energies. Hence, it was chosen as the event generator for this study as well. The kinetic energy dependence of $p_0$ for the antideuteron production using {\tt EPOS-LHC} was described by the following parametrization:
\begin{equation}
p_0\left( T \right) = \frac{A}{1 + \exp\left( B - \ln(T/C) \right)}
\label{eqn:coalescenceparameterization}
\end{equation}
where $T$ is the collision kinetic energy in GeV and the parameters $A$, $B$, and $C$ were determined in Ref.~\cite{Gomez-Coral:2018yuk} to be $89.6 \pm 3.0$\,MeV/$c$, $6.6 \pm 0.88$, and $0.73 \pm 0.10$, respectively. 

\subsection{Coalescence of larger antinuclei}\label{s1ss2}
In this work, the event-by-event coalescence mechanism of formation of two-particle nuclei (deuterons and antideuterons) was extended to estimate the production of larger antinuclei (${}^3\overline{\text{He}}$ and ${}^4\overline{\text{He}}$). Two simplistic scenarios were considered:  

\begin{figure*}[t]
\begin{center}
\begin{tabular}{l l}
\includegraphics[width=0.5\textwidth]{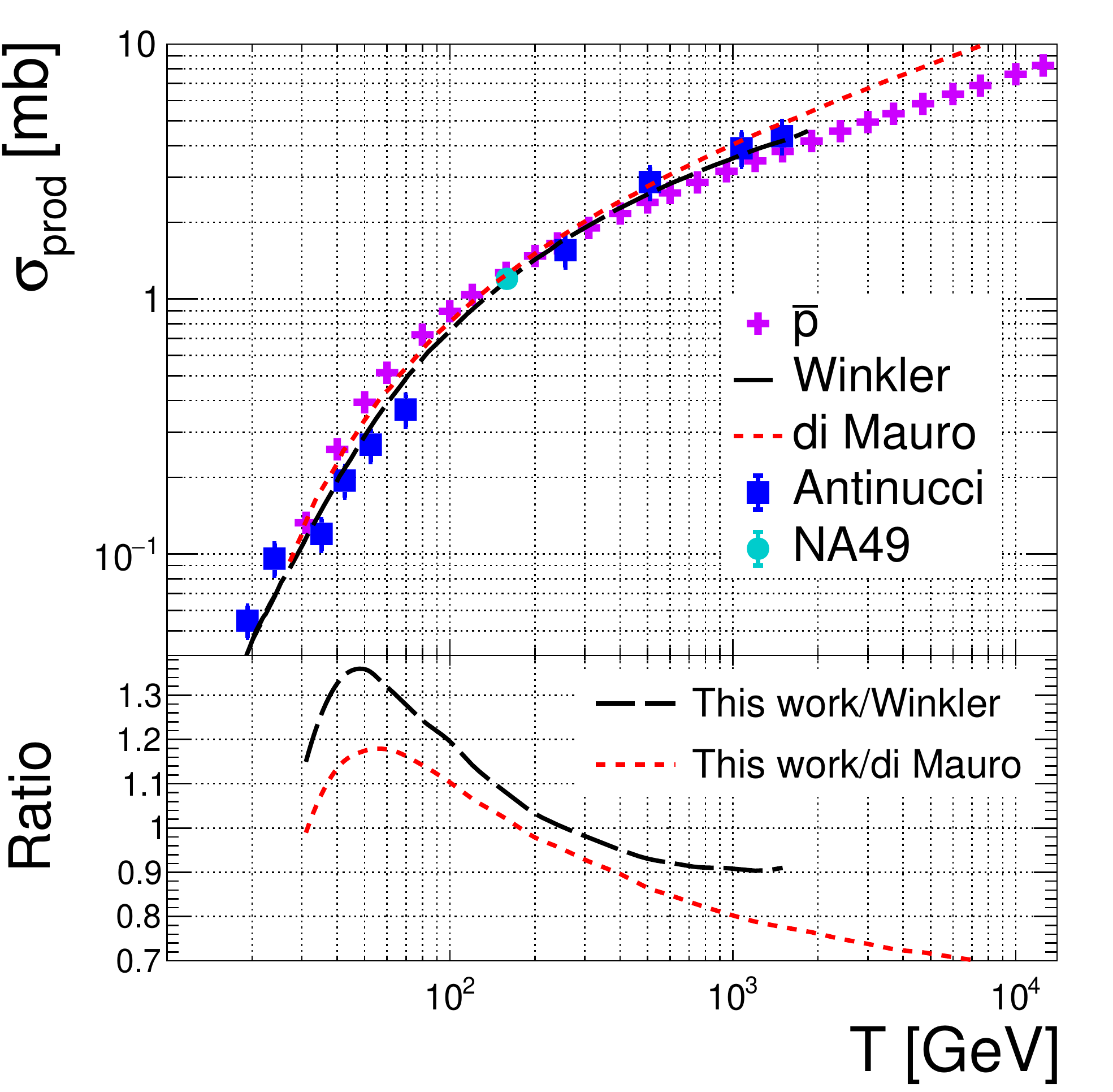} &
\includegraphics[width=0.5\textwidth]{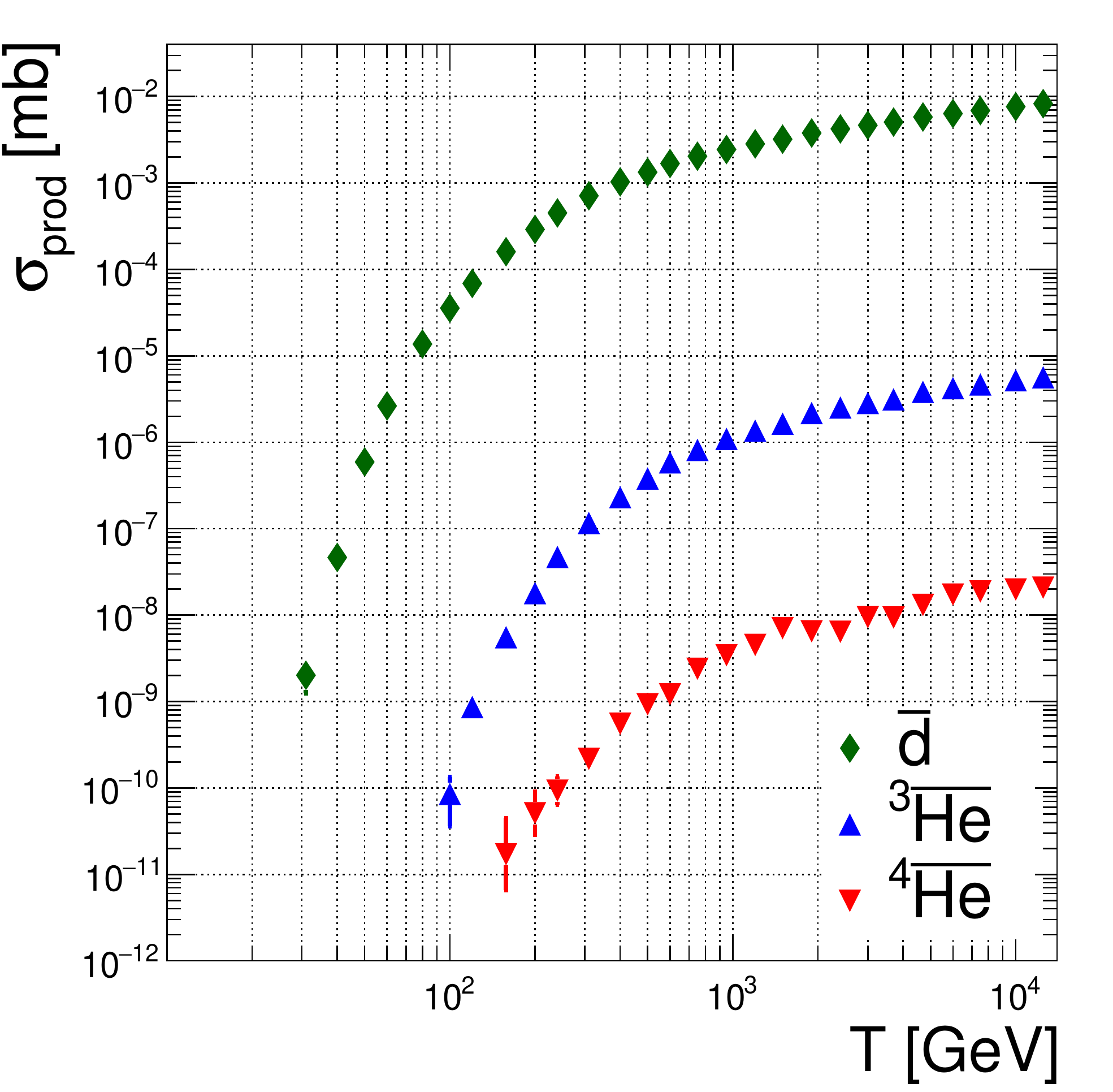}
\end{tabular}
\caption{Production cross sections calculated in this study for (left) $\overline{\text{\textit{p}}}$ and (right) heavier antinuclei in proton-proton collisions as function of collision kinetic energy $T$ [GeV] (laboratory frame), using the coalescence mechanism at 120$\%$ of $p_{0,G}$. The $\overline{\text{\textit{p}}}$ production cross sections are also compared to experimental data from Refs.~\cite{Anticic_2009, osti_4593576} and parametrizations from Winkler~\cite{winkler2017cosmic} and di Mauro~\cite{PhysRevD.90.085017, Reinert:2017aga}.}
\label{fig:pbar_Xsec}
\end{center}
\end{figure*}

\textit{i) Simultaneous coalescence}.---An $N$-particle antinucleus is formed by simultaneously coalescing $N$ antiproton and antineutrons, where each antiproton and antineutron pair has to fulfill the aforementioned coalescence condition (Eq.~\ref{eqn:coalescenceCondition}). For example, to produce ${}^4\overline{\text{He}}$, two antiprotons and two antineutrons are selected, and the coalescence condition is evaluated for the six possible particle pairs. 

\textit{ii) Iterated coalescence}.---Antiprotons or antineutrons are iteratively added to a multiantinucleon state if they fulfill the two-particle coalescence condition. For example, an antideuteron produced by the \textit{simultaneous coalescence} scenario is further evaluated for ${}^3\overline{\text{He}}$ production, by pairing it with all remaining antiprotons in that event. Similarly, ${}^3\overline{\text{He}}$ is paired with all remaining antineutrons to check for ${}^4\overline{\text{He}}$ production.

This study used both these methods. The \textit{simultaneous coalescence} method is first used to generate the initial antiparticles, and then the \textit{iterated coalescence} method is used to produce additional antiparticles.

\sisetup{tight-spacing=true}

\begin{table}[t]
\caption{Energy bins, number of generated events per bin, and antiparticles produced per event using the coalescence model at 120$\%$ of $p_{0,G}$.}
\begin{ruledtabular}
\begin{tabular}{p{1cm} p{2.0cm} c c c}

\multirow{3}{=}{Energy Bins (GeV)} & \multirow{2}{=}{\centering{Number of events (billion)}} & \multirow{2}{*}{$\overline{\text{\textit{d}}}$} & \multirow{2}{*}{${}^3\overline{\text{He}}$} & \multirow{2}{*}{${}^4\overline{\text{He}}$} \\
&  &  & \\
\cline{3-5}
&  &  \multicolumn{3}{c}{Particles produced per event}\\
\hline

	31  &   106  &  \num{6.60e-11}  &               0  &               0   \\
	40  &   843  &  \num{1.52e-09}  &               0  &               0   \\
	50  &   431  &  \num{1.92e-08}  &               0  &               0   \\
	60  &   440  &  \num{8.56e-08}  &               0  &               0   \\
	80  &   583  &  \num{4.40e-07}  &               0  &               0   \\
 100  &  1100  &  \num{1.14e-06}  &  \num{2.73e-12}  &               0   \\
 120  &  1133  &  \num{2.18e-06}  &  \num{2.74e-11}  &               0   \\
 158  &  1865  &  \num{5.02e-06}  &  \num{1.74e-10}  &  \num{5.36e-13}   \\
 200  &  1895  &  \num{8.99e-06}  &  \num{5.57e-10}  &  \num{1.58e-12}   \\
 240  &  2441  &  \num{1.39e-05}  &  \num{1.45e-09}  &  \num{2.87e-12}   \\
 310  &  2461  &  \num{2.16e-05}  &  \num{3.53e-09}  &  \num{6.09e-12}   \\
 400  &  2583  &  \num{3.09e-05}  &  \num{6.95e-09}  &  \num{1.66e-11}   \\
 500  &  1994  &  \num{3.96e-05}  &  \num{1.13e-08}  &  \num{2.76e-11}   \\
 600  &  1147  &  \num{4.95e-05}  &  \num{1.72e-08}  &  \num{3.57e-11}   \\
 750  &  1017  &  \num{5.94e-05}  &  \num{2.37e-08}  &  \num{6.98e-11}   \\
 950  &  1001  &  \num{7.00e-05}  &  \num{3.14e-08}  &  \num{9.88e-11}   \\
1200  &   795  &  \num{8.04e-05}  &  \num{3.86e-08}  &  \num{1.28e-10}   \\
1500  &   424  &  \num{9.02e-05}  &  \num{4.60e-08}  &  \num{1.98e-10}   \\
1900  &   566  &  \num{1.05e-04}  &  \num{6.04e-08}  &  \num{1.80e-10}   \\
2400  &   296  &  \num{1.16e-04}  &  \num{6.89e-08}  &  \num{1.75e-10}   \\
3000  &   258  &  \num{1.26e-04}  &  \num{7.68e-08}  &  \num{2.55e-10}   \\
3700  &   268  &  \num{1.35e-04}  &  \num{8.37e-08}  &  \num{2.50e-10}   \\
4700  &   333  &  \num{1.52e-04}  &  \num{1.01e-07}  &  \num{3.45e-10}   \\
6000  &   212  &  \num{1.64e-04}  &  \num{1.09e-07}  &  \num{4.47e-10}   \\
7500  &   275  &  \num{1.76e-04}  &  \num{1.19e-07}  &  \num{4.79e-10}   \\
10000  &   257  &  \num{1.92e-04}  &  \num{1.31e-07}  &  \num{4.96e-10}   \\
12500  &   308  &  \num{2.04e-04}  &  \num{1.39e-07}  &  \num{5.15e-10}   \\

\end{tabular}
\end{ruledtabular}
\label{s1:tab2}
\end{table}

\begin{figure*}[t]
\begin{center}
\begin{tabular}{l l}
\includegraphics[width=0.5\textwidth]{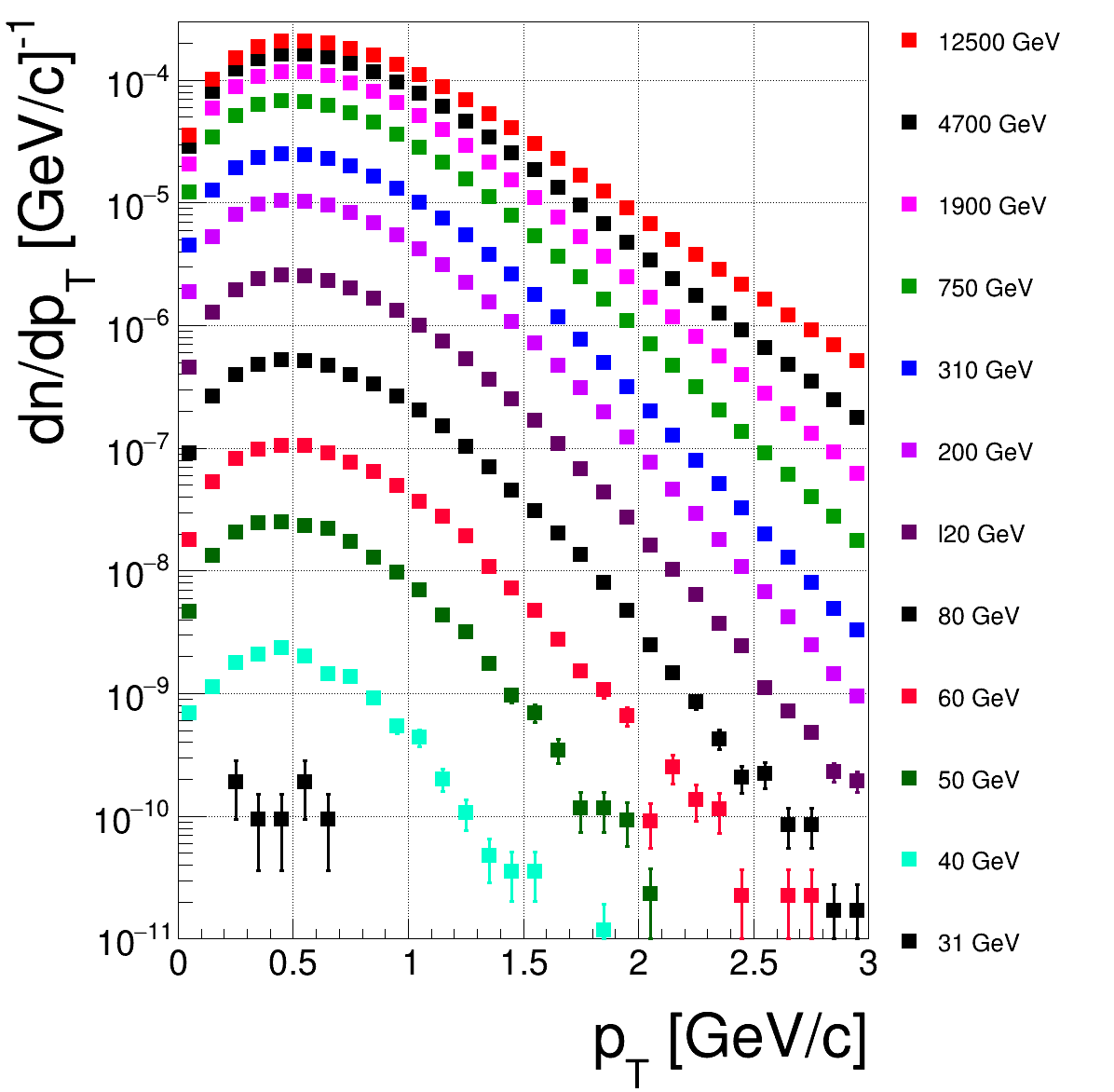} &
\includegraphics[width=0.5\textwidth]{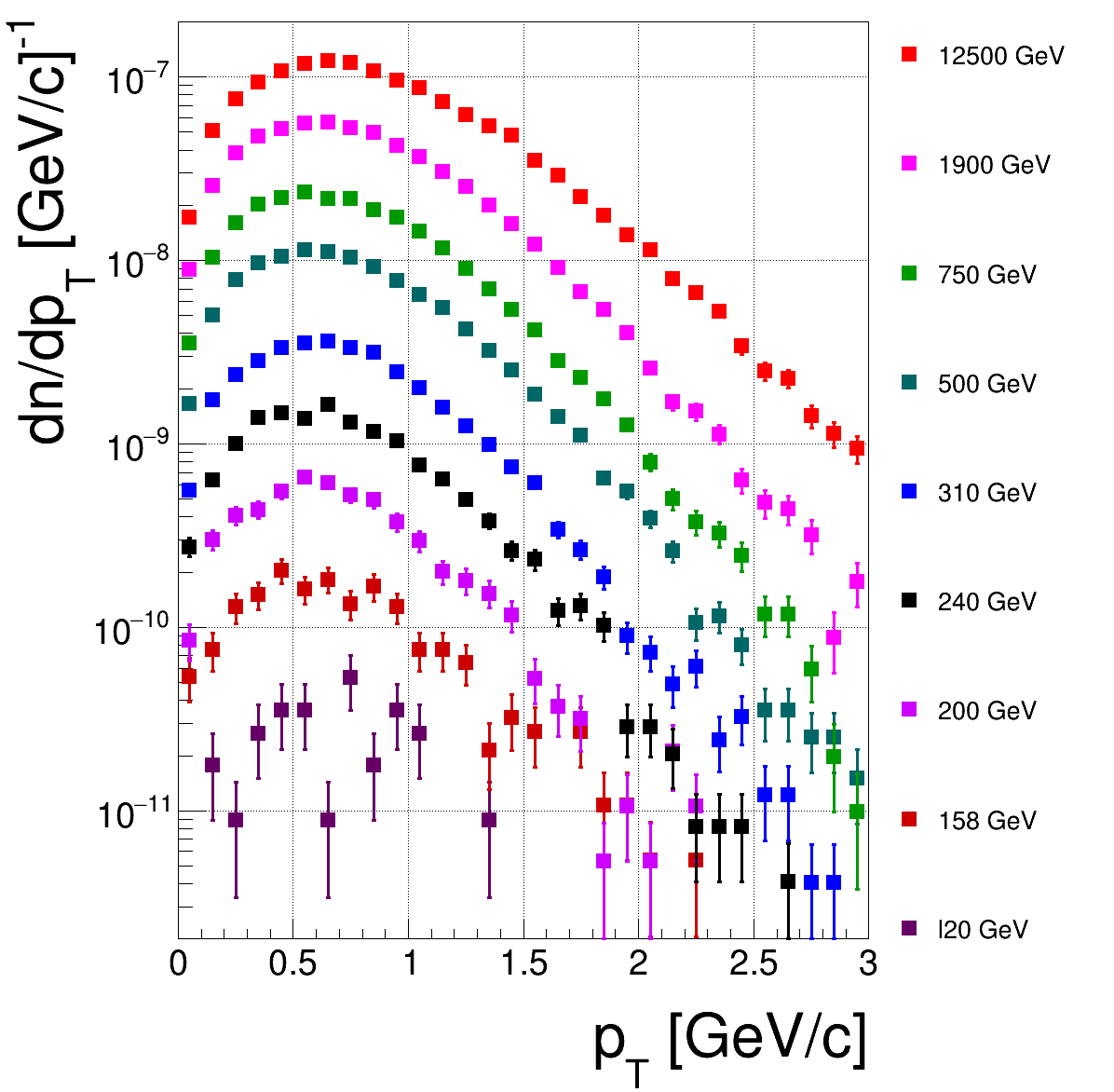}
\end{tabular}
\caption{Antinuclei spectra produced via the coalescence mechanism as a function of its transverse momentum $p_T$ (GeV/$c$) are plotted for selected CR energies (laboratory frame), using the coalescence mechanism at 120$\%$ of $p_{0,G}$: (left) $\overline{\text{\textit{d}}}$ spectra and (right) ${}^3\overline{\text{He}}$ spectra. }
\label{fig:hist_dNdpT}
\end{center}
\end{figure*}

The $p_0$ parametrization used in this study was obtained from the study of antideuteron formation by Gomez \textit{et al.}~\cite{Gomez-Coral:2018yuk}. This was done to test the hypothesis that a single parameter can accurately describe the formation of heavier antinuclei. This is supported by previous work on the analytical coalescence model, where it was shown that the two-particle coalescence parameter calculated from $\overline{\text{\textit{d}}}$ production could be scaled to correctly predict the ${}^3\overline{\text{He}}$ production as well~\cite{Acharya:2017fvb, 2018arXiv180808961P, Korsmeier:2017xzj, Blum:2017qnn}.

For a systematic study of the dependence of antinuclei production on $p_0$, seven different values of $p_0$ for each collision energy were used in this work. These seven values include an initial value of $p_0$ specific to that kinetic energy, from the $\overline{\text{\textit{d}}}$ parametrization developed by Gomez \textit{et al.} This parametrization is shown in Eq.~\ref{eqn:coalescenceparameterization}. For the rest of this study, this initial value is referred to as $p_{0,G}$. The remaining six values are 70$\%$, 80$\%$, 90$\%$, 110$\%$, 120$\%$, and 130$\%$ of $p_{0,G}$. 

Proton-proton interactions were simulated at 27 collision energy values, in logarithmic bins between 31\,GeV and 12.5\,TeV in the laboratory frame. For each collision, a projectile proton moving with the selected energy was collided head on with a stationary proton target. These simulated collisions mimic the interaction of cosmic rays with the interstellar matter. The afterburner was used to implement the coalescence conditions to generate light antinuclei.

The number of collisions simulated in each bin was motivated by that bin's relative contribution to the overall source term of the produced antiparticles (see the discussion in Sec.~\ref{s3}). It was estimated that the contribution to the $\overline{\text{\textit{d}}}$, ${}^3\overline{\text{He}}$, and ${}^4\overline{\text{He}}$ source terms are the largest from the bins at 158, 310, and 400\,GeV, respectively. Hence, the bins from 158-500\,GeV have the most number of simulated events to get the best estimates of the production cross sections. Further, since $p$-$p$ collisions contribute 60\%-70\% of the total antinuclei source terms~\cite{GomezCoral:2673048, Korsmeier:2017xzj}, only those were simulated for this study. The remaining contributions ($p$-He, He-$p$, and He-He) were estimated by scaling the parametrization developed in Ref.~\cite{Reinert:2017aga}.

The total number of $p$-$p$ collisions simulated for each energy bin and the number of antiparticles produced per $p$--$p$ collision by the coalescence mechanism are given in Table~\ref{s1:tab2}. Figure~\ref{fig:pbar_Xsec} (right) shows the production cross section of these antiparticles as a function of collision energy (in the c.m. frame). The figure shows that the production of antiprotons increases with energy and eventually saturates at high energy (approximately 1\,TeV). Antideuteron production also increases with energy. Both ${}^3\overline{\text{He}}$ and ${}^4\overline{\text{He}}$ show a similar feature but with a higher production threshold, and their saturation occurs at progressively higher kinetic energies as well. As expected, a clear trend is observed that antinuclei production becomes rarer as the number of antinucleons in the final state increases.

Figure~\ref{fig:pbar_Xsec} (left) shows a large gap in the available $\overline{\text{\textit{p}}}$ production cross section data in the few-hundred GeV range. Similarly, there are very few data points for $\overline{\text{\textit{d}}}$ production cross sections in $p$-$p$ interactions at low collision energies, with the data point at $p_{\text{lab}} = 70$\,GeV/$c$~\cite{Abramov_Baldin1987} being followed by the next available datum at $p_{\text{lab}} = 1500$\,GeV/$c$. More experimental data in the low-energy region near the production thresholds are crucial, as this is the dominant region for the production of antinuclei in cosmic-ray interactions. Latest results from the NA61/SHINE experiment at CERN-SPS at $p_{\text{lab}} = 158$\,GeV/$c$~\cite{Aduszkiewicz_2017} and analysis of new large datasets are very important. Figure~\ref{fig:hist_dNdpT} shows the predicted production yields of $\overline{\text{\textit{d}}}$ and ${}^3\overline{\text{He}}$ as a function of transverse momentum $p_T$, for selected collision energies in the laboratory frame.

\section{Validating the multi-particle coalescence approach}\label{s2}

\subsection{Comparison with $\overline{\text{\textit{p}}}$ production data}
\begin{figure*}[t]
\begin{tabular}{l l}
\includegraphics[width=0.5\textwidth]{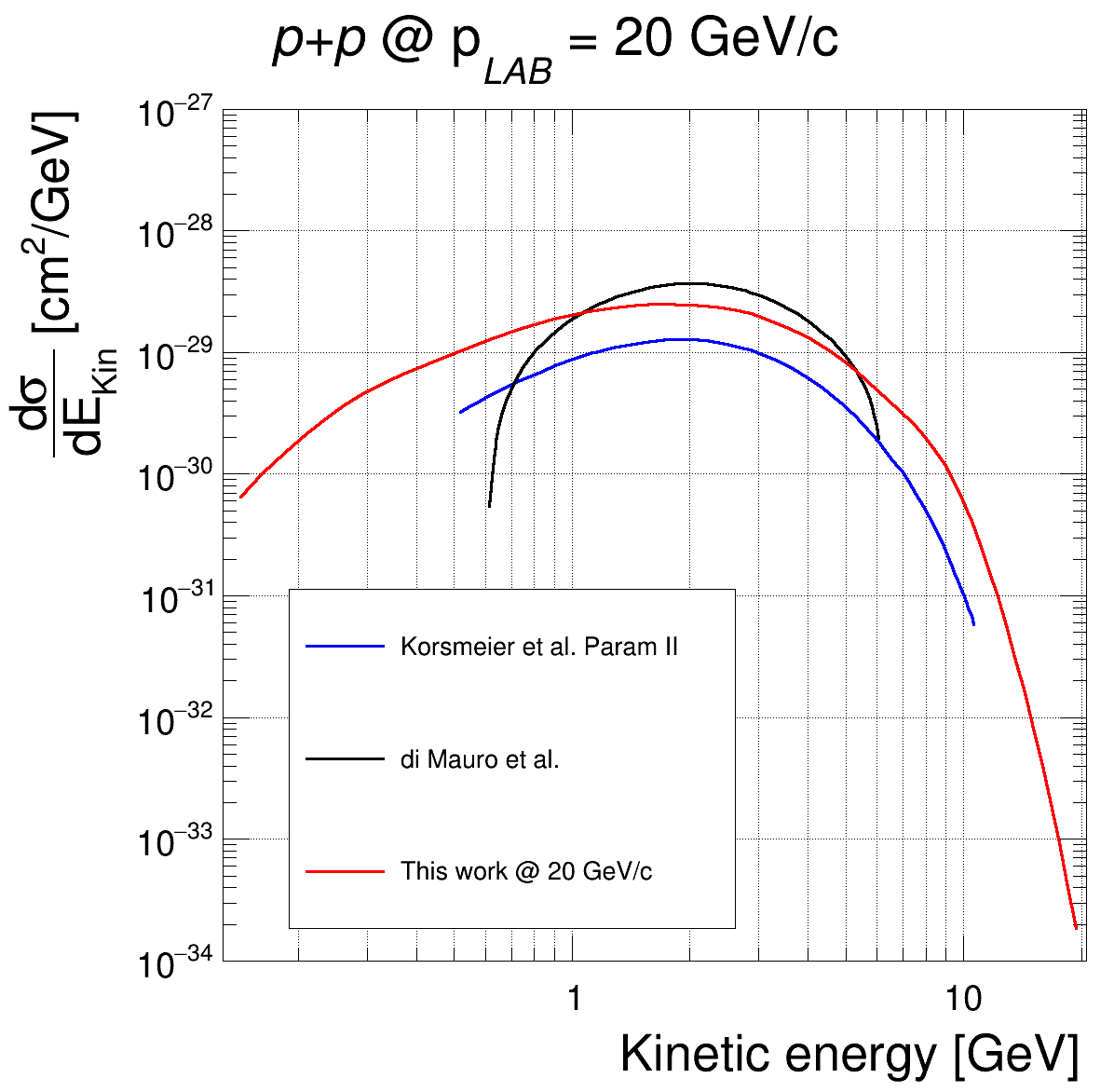}
\includegraphics[width=0.5\textwidth]{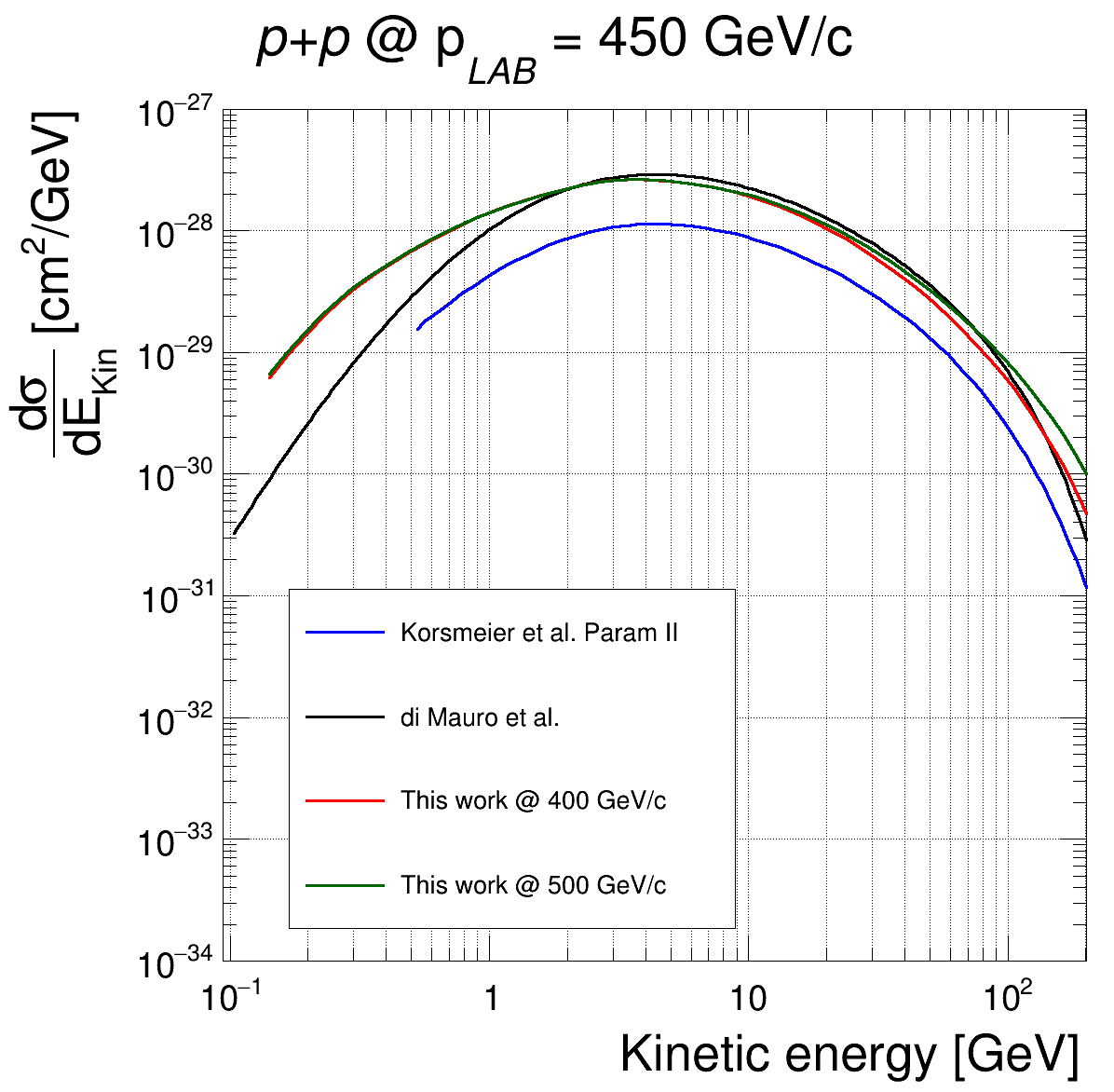}
\end{tabular}
\caption{The antiproton differential production cross section as function of kinetic energy $E_{\text{Kin}}$ for (left) $p$-$p$ at $p_{\text{lab}} = 20$\,GeV/$c$ and (right) $p$-$p$ at $p_{\text{lab}} = 450$\,GeV/$c$. The results are compared to parametrization from di Mauro \textit{et al.}~\cite{PhysRevD.90.085017} and Korsmeier \textit{et al.}~\cite{Korsmeier:2018gcy}. The parametrization from Korsmeier \textit{et al.} does not include the contribution from decay of antineutrons.}
\label{fig:pp20_450}
\end{figure*}

The predicted antinuclei fluxes from cosmic-ray propagation models are highly correlated with antiproton production in proton-proton interactions. Hence, it is important to get the correct antiproton production cross section. The $\overline{\text{\textit{p}}}$ total production cross section as predicted from this simulation was compared with data at different collision energies~\cite{Anticic_2009, osti_4593576} in Figure~\ref{fig:pbar_Xsec} (left). This {\tt EPOS-LHC}-based simulation is compatible with the data within the statistical uncertainties. Next, the $\overline{\text{\textit{p}}}$ differential production cross section as function of kinetic energy was compared to the latest parametrization at different collision energies. Figure~\ref{fig:pp20_450} shows the comparison for collisions at 20 and 450\,GeV/$c$. The parametrization from Korsmeier \textit{et al.}~\cite{Korsmeier:2018gcy} shows only the primary $\overline{\text{\textit{p}}}$ cross section and does not include the contribution from $\overline{\text{\textit{n}}}$ decay; hence, it is lower than the di Mauro \textit{et al.} parametrization~\cite{PhysRevD.90.085017} used by Poulin \textit{et al.}~\cite{2018arXiv180808961P} by a factor of 2. Taking this factor into account, the cross section predicted by {\tt EPOS-LHC} in this work is in very good agreement with Ref.~\cite{Korsmeier:2018gcy}. The agreement with parametrization from di Mauro \textit{et al.} is poor for 20\,GeV/$c$ interactions but gets much better at 450\,GeV/$c$. The agreement at very low kinetic energies is especially poor.

\begin{figure*}[t]
\begin{tabular}{l l}
\includegraphics[width=0.5\textwidth]{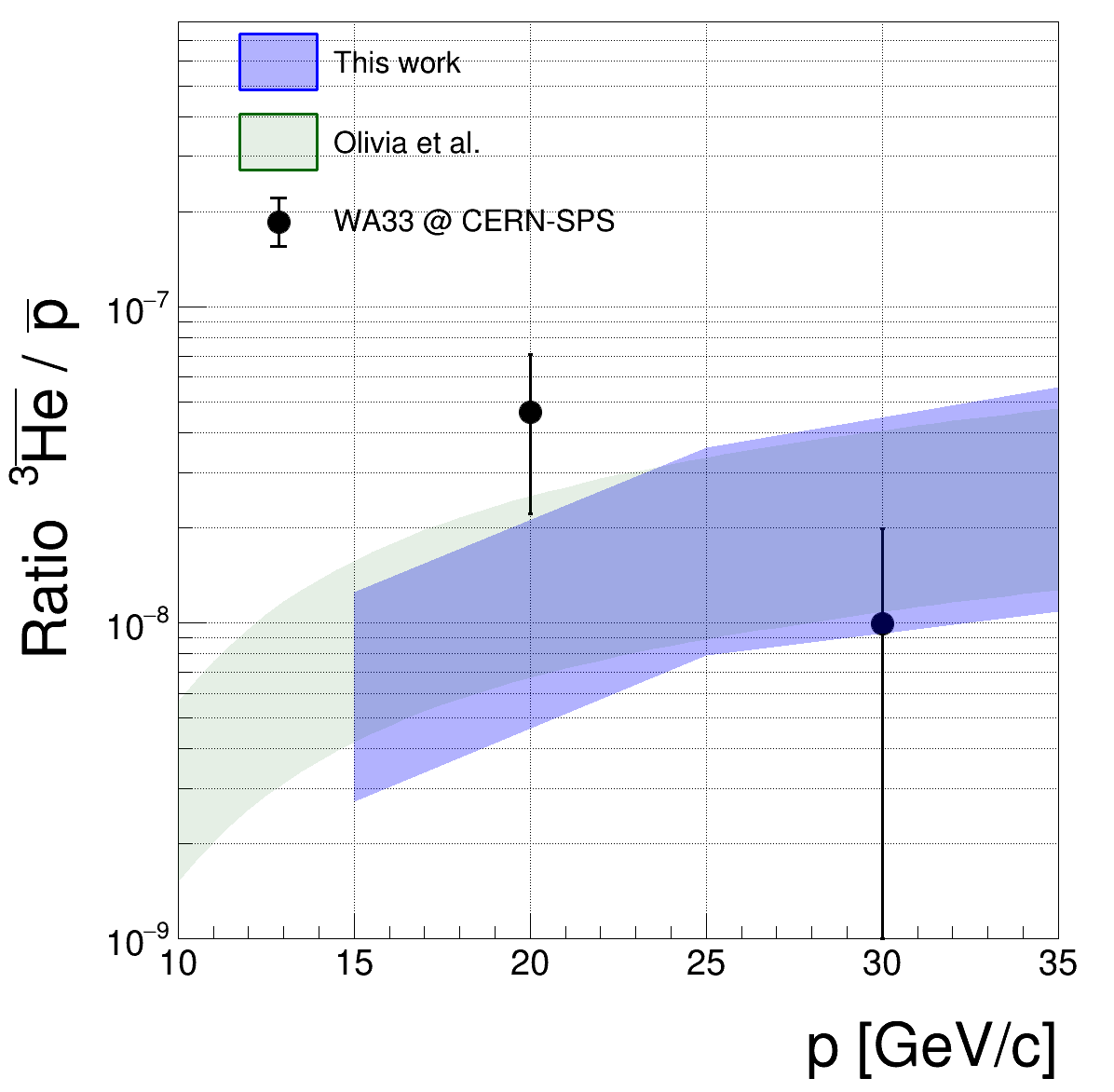}
\includegraphics[width=0.5\textwidth]{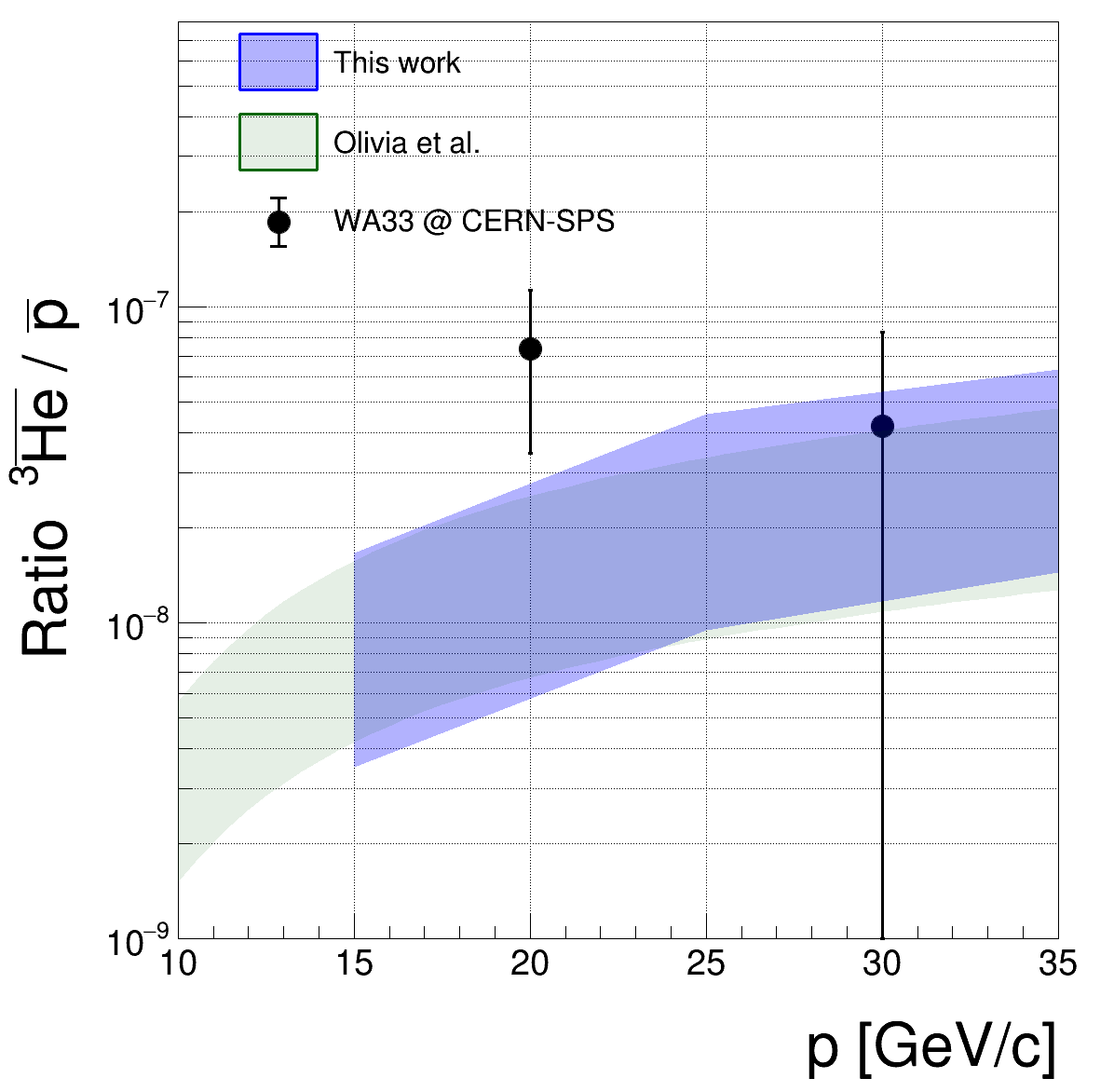}
\end{tabular}
\caption{The invariant production cross section ratio ${}^3\overline{\text{He}}$/$\overline{\text{\textit{p}}}$ as function of momentum $p$ [GeV/$c$] in the laboratory frame for (left) $p$-Be at $p_{\text{lab}} = 200$\,GeV/$c$ and (right) $p$--Al at $p_{\text{lab}} = 200$\,GeV/$c$. The uncertainty bands for this work were estimated by varying the coalescence parameter from $p_{0,G}$ (59\,MeV/$c$) to 130$\%$ of $p_{0,G}$ (77\,MeV/$c$).}
\label{s3:fig_pBe_pAl}
\end{figure*}

\subsection{Validation with $\overline{\text{\textit{d}}}$, $\overline{\text{\textit{t}}}$ and ${}^3\overline{\text{He}}$ production data}
Because of a lack of light-antinuclei production data for $p$-$p$ collisions at low energies near the production threshold, a direct comparison with the predictions of the multiparticle coalescence model is not possible. However, comparison with $p$-A collisions (where A is a light antinucleus) can produce a target-independent parametrization for the production of light antinuclei. $\overline{\text{\textit{t}}}$/$\overline{\text{\textit{p}}}$ and ${}^3\overline{\text{He}}$/$\overline{\text{\textit{p}}}$ ratios have been measured in $p$-Al and $p$-Be collisions at beam momentum of 200\,GeV/$c$~\cite{BOZZOLI1978317, BAKER1974303, AlbertoOliva1707.06918}. The predictions of this model are compared to data in Figure~\ref{s3:fig_pBe_pAl}. The uncertainty bands were estimated by varying the coalescence parameter from $p_{0,G}$ (59\,MeV/$c$) to 130$\%$ of $p_{0,G}$ (77\,MeV/$c$). In magnitude and shape, it nearly overlaps with the uncertainty band from the analytical model~\cite{AlbertoOliva1707.06918} and is in good agreement with the data.

\begin{figure*}[t]
\begin{tabular}{l l}
\includegraphics[width=0.5\textwidth]{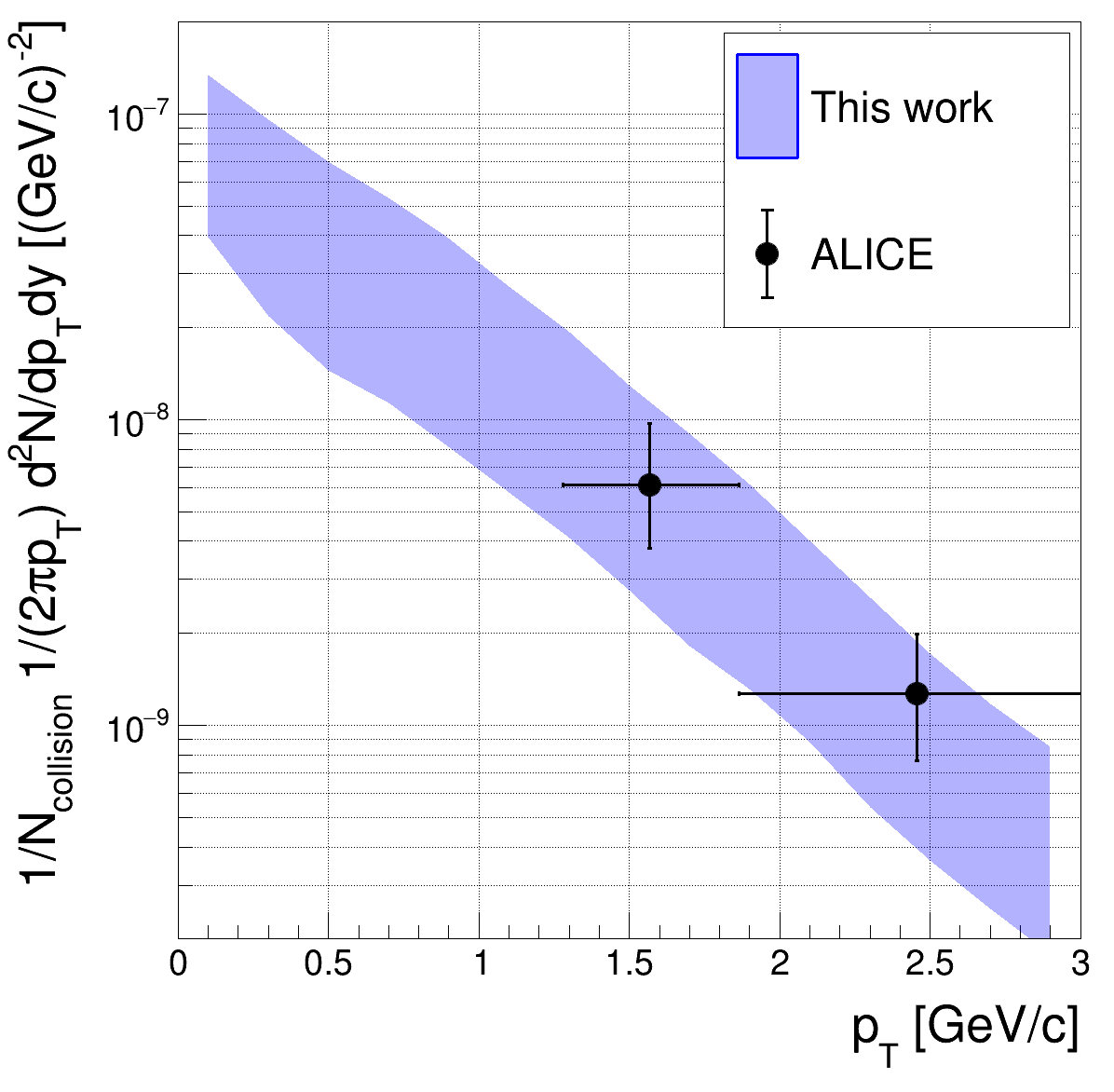}
\includegraphics[width=0.5\textwidth]{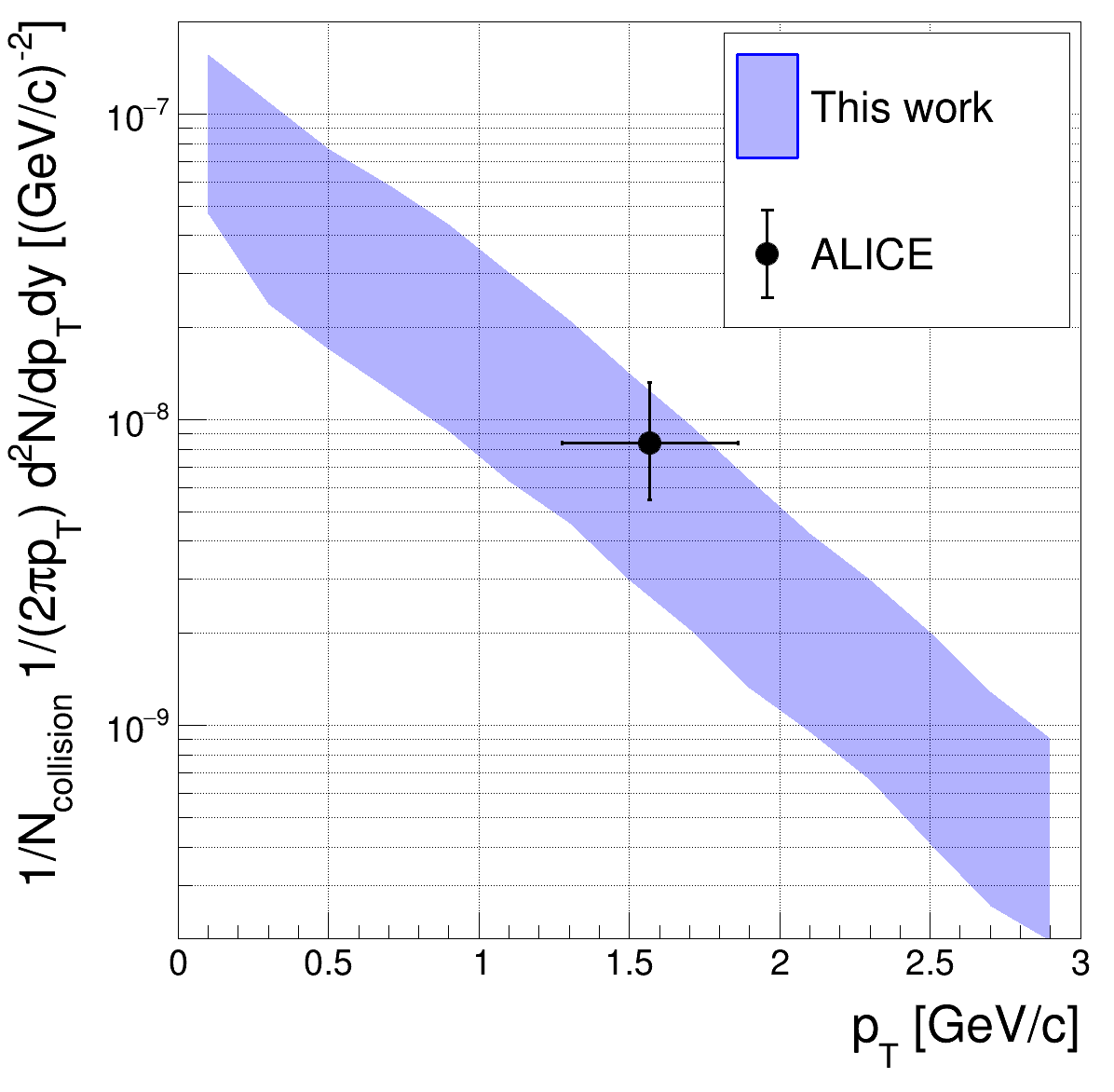}
\end{tabular}
\caption{Number density of (left) ${}^3\overline{\text{He}}$ and (right) $\overline{\text{\textit{t}}}$ production from coalescence mechanism for $p$-$p$ interactions at $\sqrt{s}$ = $7$\,TeV, along with ALICE results from Ref.~\cite{AlicePhysRevC.97.024615}. The uncertainty bands were estimated by varying the coalescence parameter from $p_{0,G}$ (90\,MeV/$c$) to 130$\%$ of $p_{0,G}$ (116\,MeV/$c$).}
\label{s3:fig1}
\end{figure*}

Moreover, proton-proton collisions were simulated at $\sqrt{s}$ = $7$\,TeV, with the aim of comparing the coalescence scheme with the latest ${}^3\overline{\text{He}}$ production data from ALICE~\cite{AlicePhysRevC.97.024615}. The coalescence parameter $p_0$ was again varied from $p_{0,G}$ (90\,MeV/$c$) to 130$\%$ of $p_{0,G}$ (116\,MeV/$c$), to simulate an uncertainty band. The results are shown in Figure~\ref{s3:fig1}. The ${}^3\overline{\text{He}}$ and $\overline{\text{\textit{t}}}$ production yields from ALICE are shown to be within 10$\%$--30$\%$ of the yield predicted by using $p_{0,G}$. Moreover, as found by ALICE and in Ref.~\cite{Ding_2019}, the simulation also shows no measurable asymmetry in antitriton and antihelium production at very high energies.

\begin{figure*}[t]
\begin{tabular}{l l}
	\includegraphics[width=0.50\textwidth]{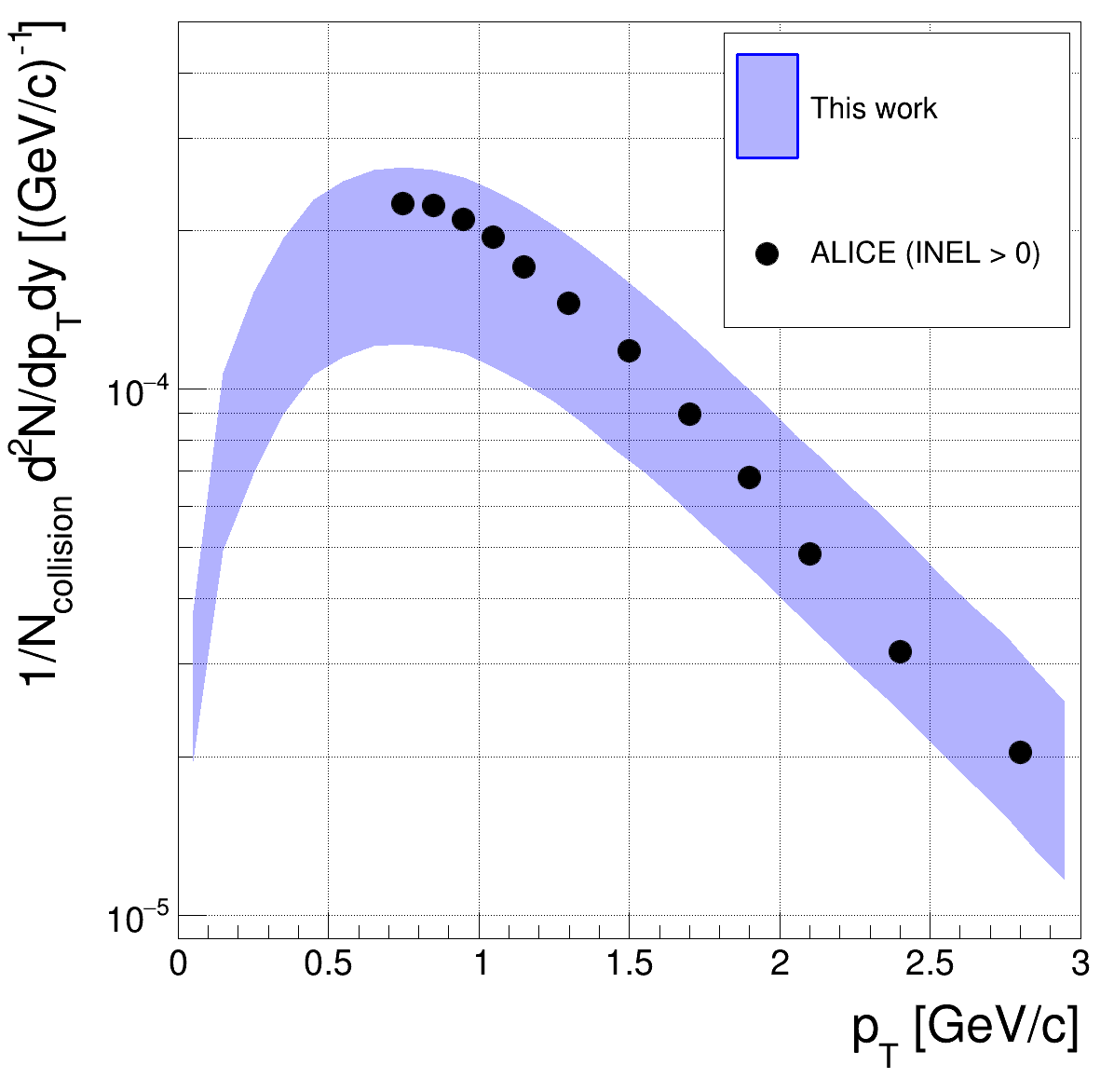}
	\includegraphics[width=0.50\textwidth]{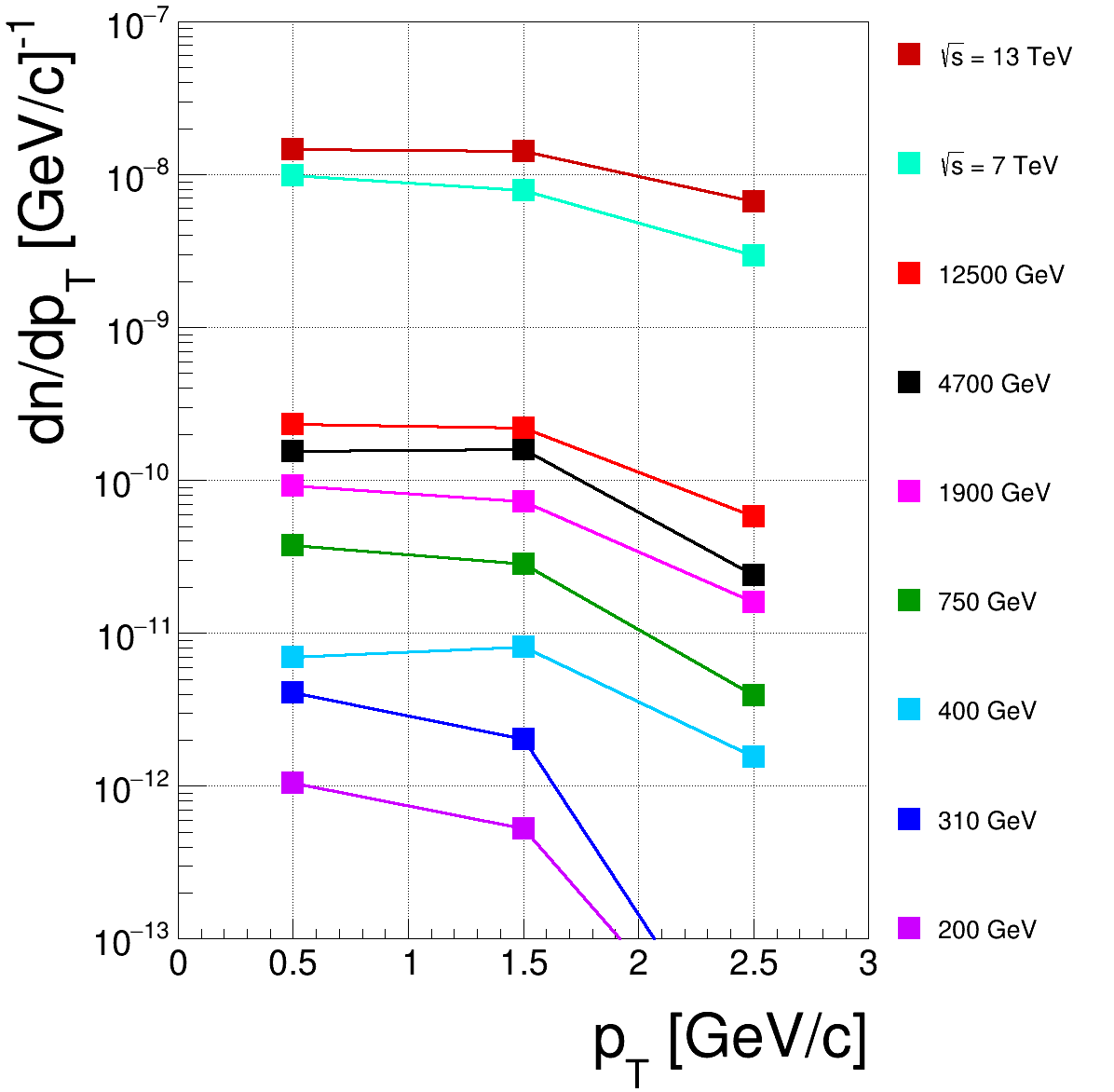}
\end{tabular}
\caption{Left: number density of $\overline{\text{\textit{d}}}$ production from coalescence mechanism for $p$-$p$ interactions at $\sqrt{s}$ = $13$\,TeV, along with ALICE results from Ref.~\cite{collaboration2020antideuteron}. The uncertainty band was estimated by varying the coalescence parameter from $p_{0,G}$ (90\,MeV/$c$) to 130$\%$ of $p_{0,G}$ (116\,MeV/$c$). Right: the predicted differential yield of ${}^4\overline{\text{He}}$ as a function of $p_{T}$ in $p$-$p$ interactions at different collision energies, using the coalescence mechanism at 120$\%$ of $p_{0,G}$.}
\label{fig:alice_13_tev}
\end{figure*}

Since the publication of $\overline{\text{\textit{d}}}$ parametrization in Ref.~\cite{Gomez-Coral:2018yuk}, new data for $\overline{\text{\textit{d}}}$ production at $\sqrt{s}$ = $13$\,TeV have been published by ALICE~\cite{collaboration2020antideuteron}. Comparison of the data with the predictions from the coalescence model is shown in Figure~\ref{fig:alice_13_tev} (left). Once again, the uncertainty band was estimated by varying the coalescence parameter from $p_{0,G}$ (90\,MeV/$c$) to 130$\%$ of $p_{0,G}$ (116\,MeV/$c$). The $\overline{\text{\textit{d}}}$ production yield from ALICE is shown to be within 10$\%$--20$\%$ of the yield predicted by using $p_{0,G}$ as the coalescence momentum.

As discussed in Sec.~\ref{s1ss2}, the production cross sections for $\overline{\text{\textit{d}}}$, ${}^3\overline{\text{He}}$, and ${}^4\overline{\text{He}}$ at each collision energy were estimated at seven different values of the coalescence parameter $p_0$. The parametrization in Eq.~\ref{eqn:coalescenceparameterization} was used to get the initial value of $p_0$ (i.e., $p_{0,G}$), and the other $p_0$ values were used to estimate the uncertainty bands. However, as shown above, using $p_{0,G}$ as the coalescence parameter underpredicted the ${}^3\overline{\text{He}}$ production cross sections by 10$\%$--20$\%$ at both high-energy and low-energy interactions. Taking this finding into consideration, the subsequent ${}^3\overline{\text{He}}$ and ${}^4\overline{\text{He}}$ production cross sections and the cosmic-ray flux discussion are shown with an uncertainty band, with the lower edge corresponding to $p_{0,G}$ and the upper edge corresponding to 130$\%$ of $p_{0,G}$.

The uncertainties in the $\overline{\text{\textit{d}}}$ parametrization from Ref.~\cite{Gomez-Coral:2018yuk} are also similar in magnitude, especially in the low-energy region (collision kinetic energy of approximately 158\,GeV) relevant for $\overline{\text{\textit{d}}}$ production in cosmic-ray interactions. Along with the comparison to data at $\sqrt{s}$ = $13$\,TeV as discussed above, and to be consistent with the other antinuclei, a similar uncertainty band (from $p_{0,G}$ to 130$\%$ of $p_{0,G}$) was chosen for $\overline{\text{\textit{d}}}$ as well. This effectively increased the value of parameter $A$ in Eq.~\ref{eqn:coalescenceparameterization} by 15$\%$, from 90 to 103\,MeV/$c$.

This study was able to successfully simulate enough $p$--$p$ collisions to be able to produce reasonable ${}^4\overline{\text{He}}$ spectra. Figure~\ref{fig:alice_13_tev} (right) shows the ${}^4\overline{\text{He}}$ production yield as a function of $p_{T}$ predicted by this study at different collision energies, including at $\sqrt{s}$ = $7$ and $13$\,TeV. ALICE has published results of ${}^4\overline{\text{He}}$ production in Pb-Pb collisions at $\sqrt{s}$ = $2.76$\,TeV~\cite{Acharya_2018} and ${}^4\overline{\text{He}}$ production upper limit in $p$-Pb collisions at $\sqrt{s}$ = $5.02$\,TeV~\cite{Acharya_2020}. However, due to the large difference in system size, these results cannot be used to validate the predictions of the $p$-$p$ MC simulations. Measuring the antideuteron and antihelium production at LHC energies is very useful to validate various formation models. However, cosmic rays at LHC energies are extremely rare, with most cosmic-ray protons having an energy of only a few GeV. Using collision systems with energies closer to the production threshold of light antinuclei is necessary to understand their production in the Galaxy~\cite{doetinchem2020cosmicray}.

\begin{figure*}[t]
\begin{tabular}{l l}
\includegraphics[width=0.50\textwidth]{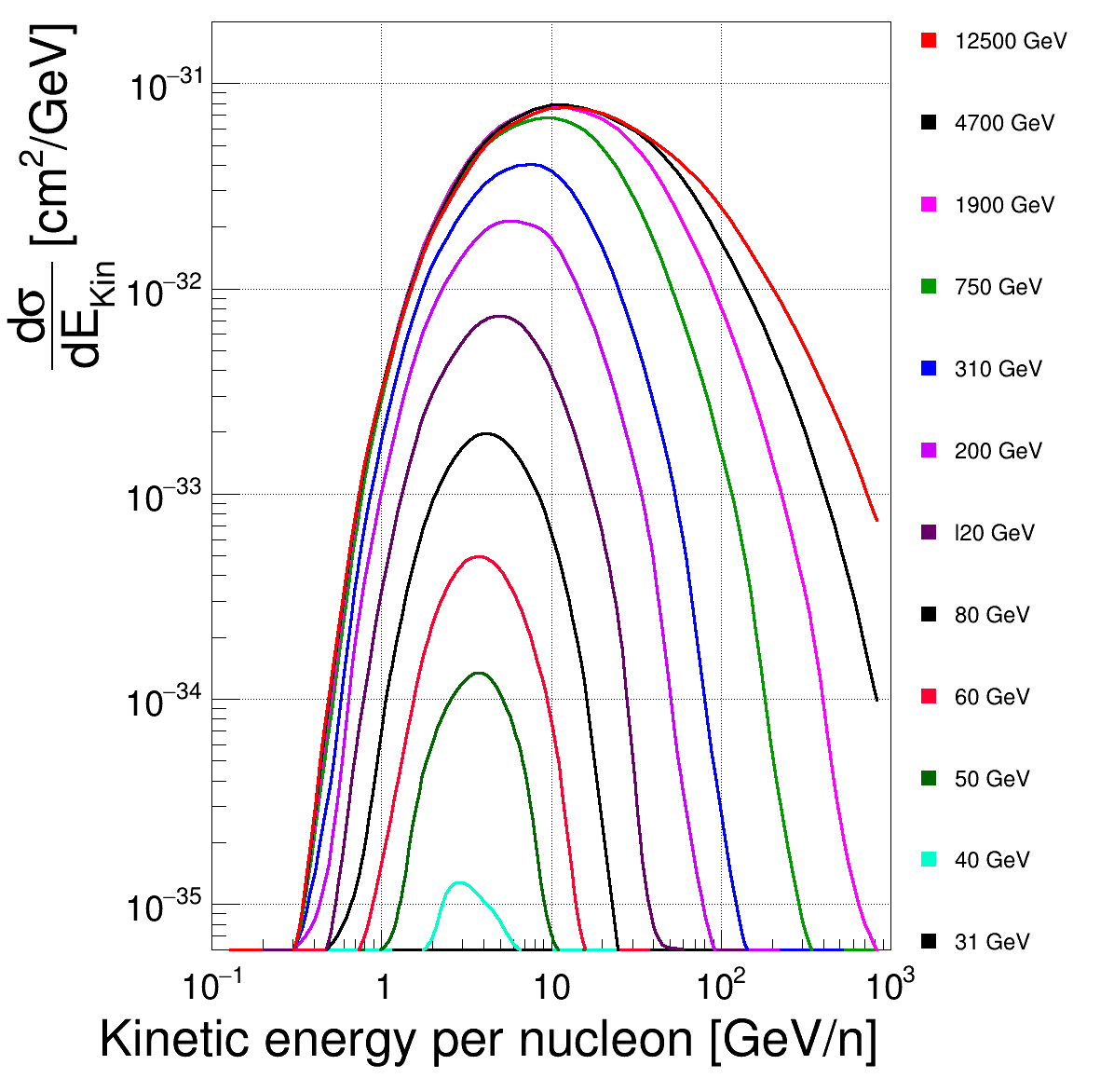}
\includegraphics[width=0.50\textwidth]{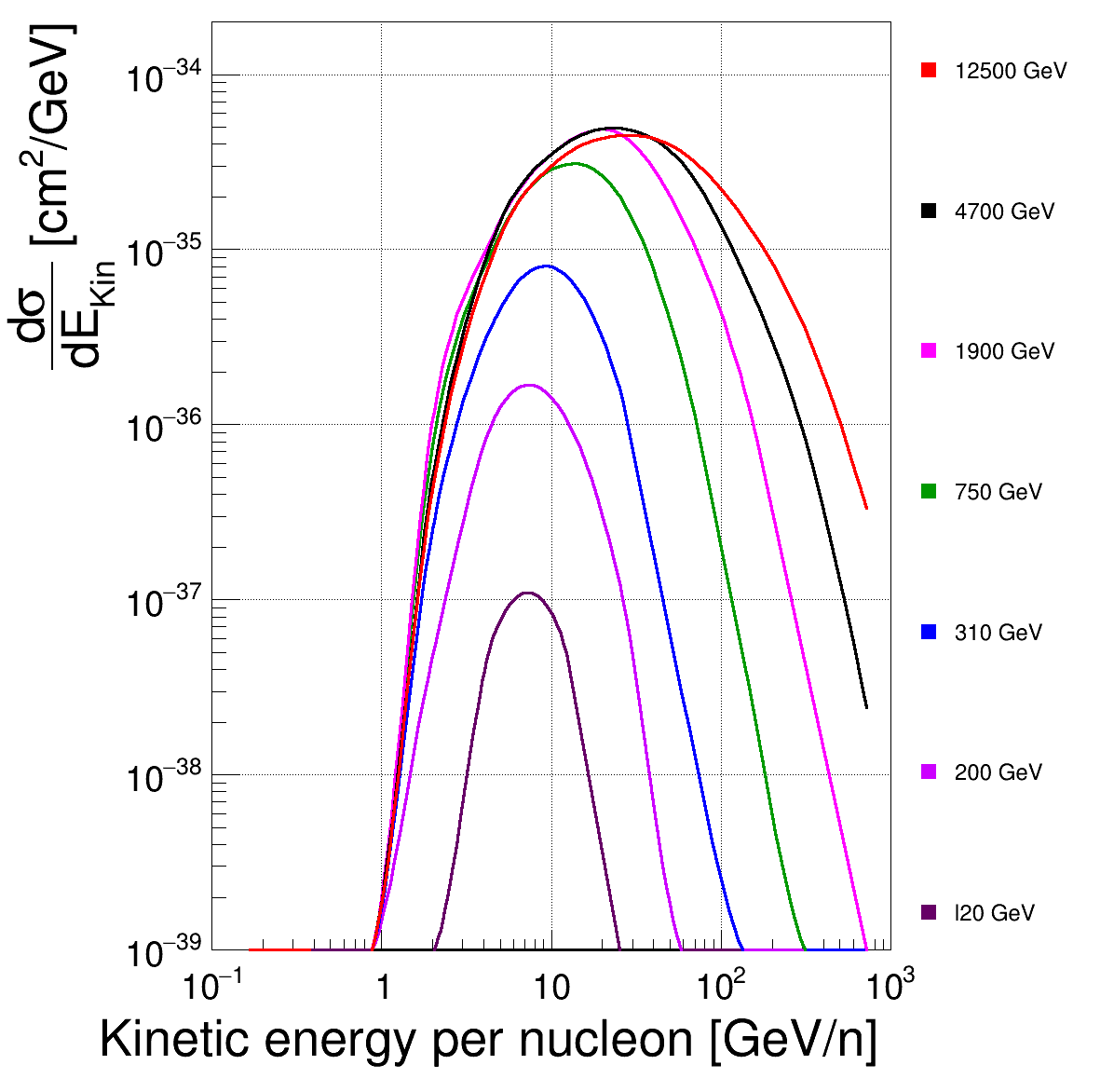}
\end{tabular}
\caption{Differential production cross section $(\text{cm}^{2}/\text{GeV})$ for (left) $\overline{\text{\textit{d}}}$ and (right) ${}^3\overline{\text{He}}$ as function of kinetic energy per nucleon $E_{\text{Kin}}$ (GeV/$n$) for selected $p$-$p$ collision energies, using the coalescence mechanism at 120$\%$ of $p_{0,G}$.}
\label{fig:hist_dSigma_dKE_dbar}
\end{figure*}

\section{Propagation of Antinuclei in the Galaxy}\label{s3}

The updated $\overline{\text{\textit{d}}}$, ${}^3\overline{\text{He}}$, and ${}^4\overline{\text{He}}$ production cross sections were used to calculate the local source terms.  This was followed by the propagation of the source terms in the Galaxy. Solar modulation was applied to produce the final top-of-the-atmosphere (TOA) fluxes.

The standard technique to estimate the antinuclei production in the interactions of cosmic rays with interstellar gas is by scaling the parametrization of $\bar{p}$ production cross section data from experiments. In this work, the production cross sections of all the antinuclei were generated using the coalescence afterburner described earlier. The $\overline{\text{\textit{d}}}$ and ${}^3\overline{\text{He}}$ differential production cross sections are shown in Figure~\ref{fig:hist_dSigma_dKE_dbar}. These cross sections were used as input for propagation in the Galaxy. For the purpose of propagation, it was assumed that over the timescale of Galactic transport, all $\overline{\text{\textit{t}}}$ decay completely into ${}^3\overline{\text{He}}$, with no change in the kinetic energy distribution. 

\begin{figure*}[t]
\begin{tabular}{l l}
\includegraphics[width=0.50\textwidth, valign=t]{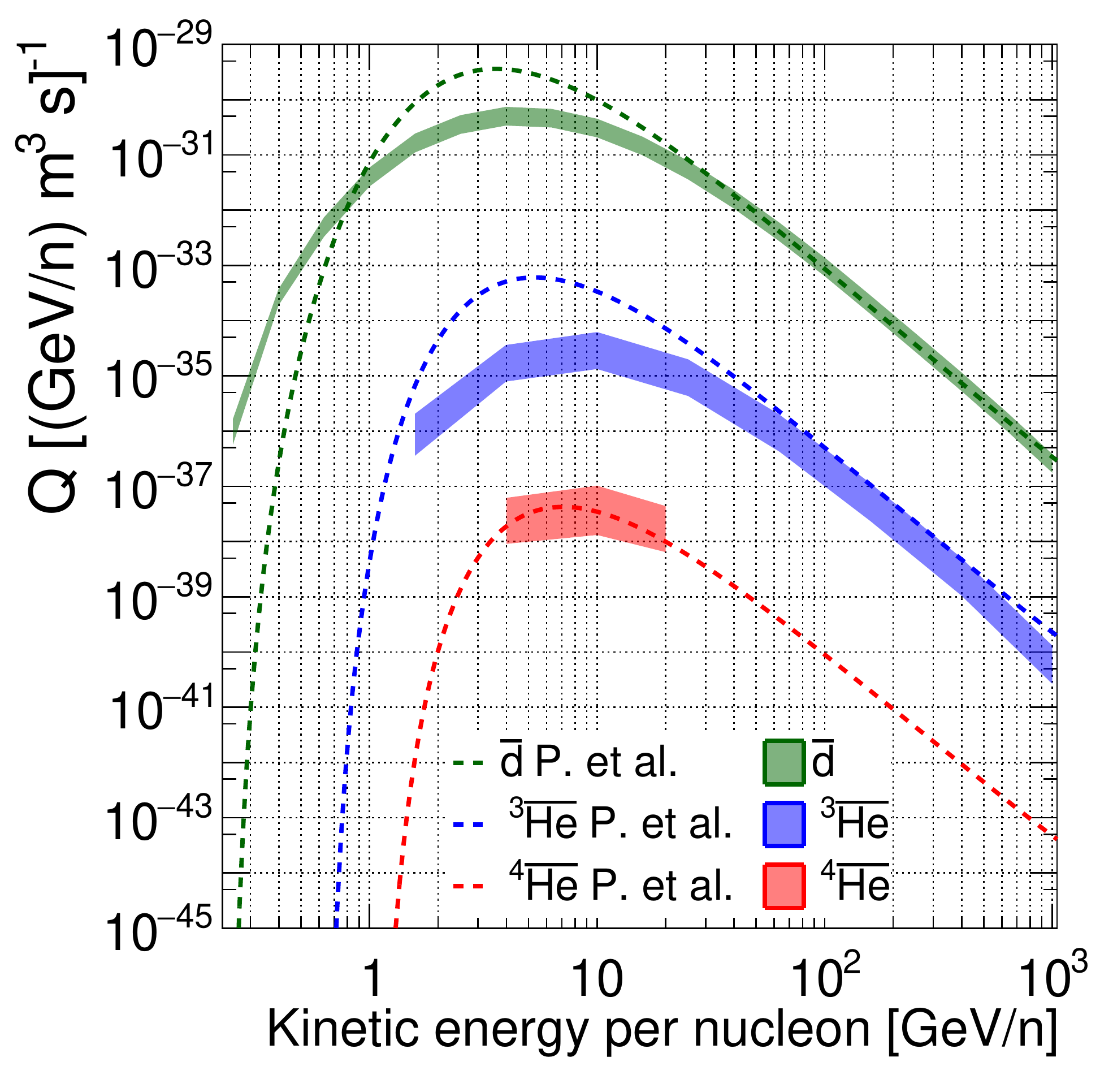}
\includegraphics[width=0.50\textwidth, valign=t]{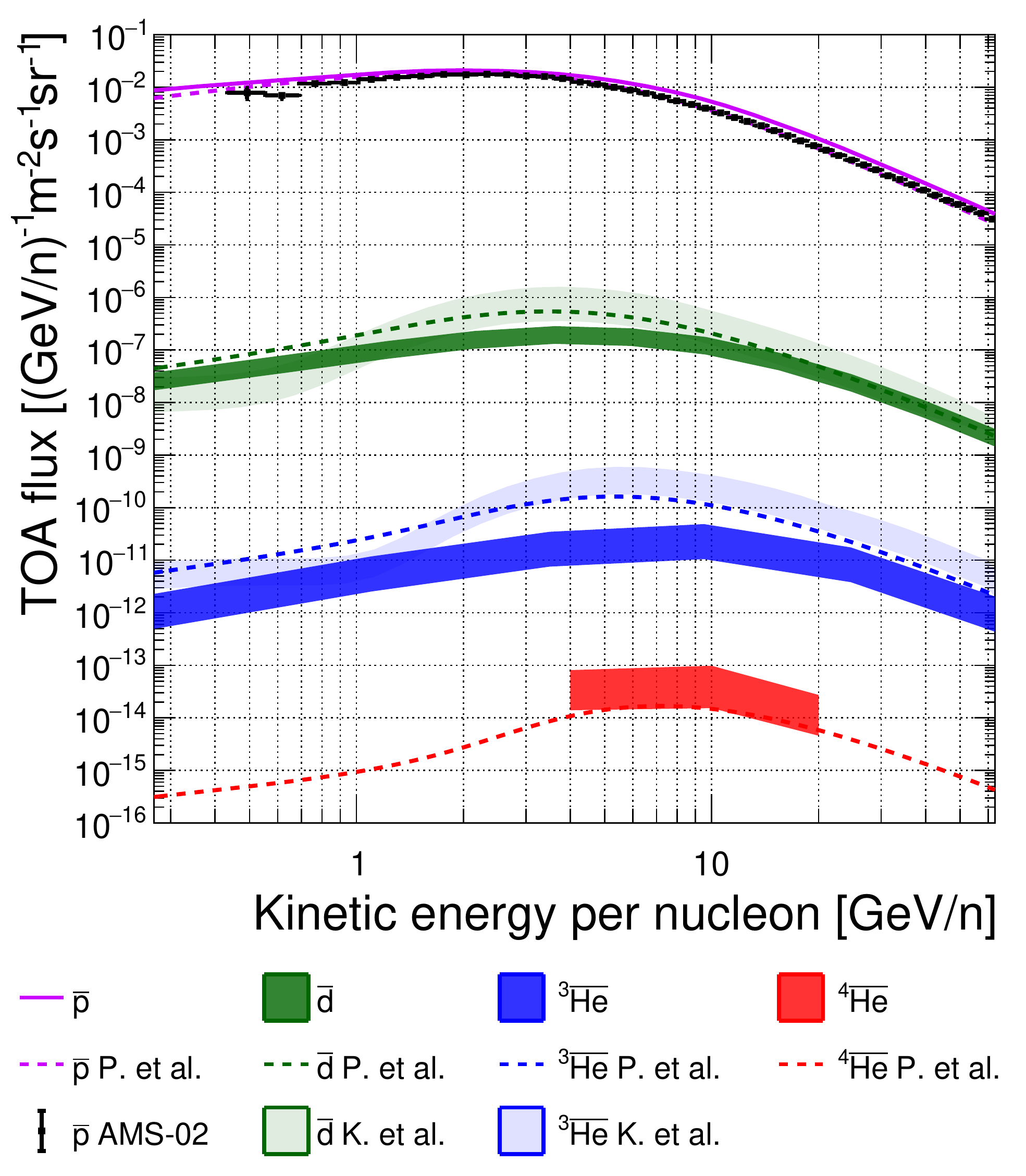}
\end{tabular}
\caption{Left: local source terms for the secondary production of $\overline{\text{\textit{d}}}$, ${}^3\overline{\text{He}}$, ${}^4\overline{\text{He}}$. The width of the bands represent uncertainty in the coalescence parameter, which was varied from $p_{0,G}$ to 130$\%$ of $p_{0,G}$. These predictions are compared to predictions by Poulin \textit{et al.}~\cite{PoulinVivianSalatiPhysRevD.99.023016} and Korsmeier \textit{et al.}~\cite{Korsmeier:2017xzj}. Right: the predicted secondary top-of-atmosphere flux for different antiparticles, propagated using the MED propagation model~\cite{Donato_2004}. The width of the bands from this work represents uncertainty in the coalescence parameter. These predictions are compared to predictions by Poulin \textit{et al.}~\cite{PoulinVivianSalatiPhysRevD.99.023016} and Korsmeier \textit{et al.}~\cite{Korsmeier:2017xzj}. The $\overline{\text{\textit{p}}}$ flux is compared to data from AMS-02~\cite{PhysRevLett.117.091103}.}
\label{fig:sourceTerm}
\end{figure*}

The flux of cosmic-ray protons $\phi_{p}(E)$ at the selected energies are obtained from the high-precision measurements by the AMS-02 Collaboration~\cite{PhysRevLett.117.091103}. The differential production cross section of an antinucleon A as a function of its kinetic energy per nucleon $E_{\text{A}}$ is obtained from the MC simulation (Figure~\ref{fig:hist_dSigma_dKE_dbar}). The local source term $\text{Q}_{\text{sec}}$ can then be calculated using~\cite{PoulinVivianSalatiPhysRevD.99.023016, PhysRevD.96.103021, 10.1140epja}
\begin{equation}
\text{Q}_{\text{sec}}(E_{\text{A}}) = 4\pi n_{\text{H}} \int_{E_{th}}^{\infty} \text{d}E \phi_{p}(E) \frac{\text{d}\sigma_{\text{A}}}{\text{d}E_{\text{A}}} (E,E_{\text{A}})
\end{equation}
where $n_H$ is the number density of hydrogen nuclei in the ISM which was set to 0.9 $\text{atoms/cm}^{\text{3}}$. The secondary antiparticle source terms as a function of the antiparticle's kinetic energy per nucleon are presented in Figure~\ref{fig:sourceTerm} (left). Both $\overline{\text{\textit{d}}}$ and ${}^3\overline{\text{He}}$ source terms are lower than the source terms predicted by Poulin \textit{et al.}~\cite{PoulinVivianSalatiPhysRevD.99.023016} by an order of magnitude in the low kinetic energy region (less than 10\,GeV). Because of low statistics (see Table~\ref{s1:tab2}), the ${}^4\overline{\text{He}}$ source term is shown only from 4--20\,GeV, where it is in agreement with Poulin \textit{et al.} As ${}^4\overline{\text{He}}$ production is extremely rare in $p$--$p$ collisions, being able to predict the ${}^4\overline{\text{He}}$ source term using MC simulations was only possible with a massive amount of computing power.

To propagate the antinuclei produced in interstellar medium, an updated semianalytical code developed in Refs.~\cite{PoulinVivianSalatiPhysRevD.99.023016, Boudaud_2015, Giesen_2015} was used. The only difference is the modification of $\overline{\text{\textit{p}}}$, $\overline{\text{\textit{d}}}$, and $\overline{\text{He}}$ production cross section tables for $p$-$p$ interactions, with tables generated using the coalescence model from this work.

The diffusion parameters used for Galactic propagation in this study were not tuned to fit the $\overline{\text{\textit{p}}}$ flux from this analysis with CR data. Instead, the parameters from Poulin \textit{et al.}~\cite{PoulinVivianSalatiPhysRevD.99.023016} were used with the MED propagation model~\cite{Donato_2004} to predict the secondary antinuclei TOA fluxes. The predicted antinuclei fluxes are shown in Figure~\ref{fig:sourceTerm} (right). The uncertainty bands shown for the fluxes from this work are due to the uncertainty in the coalescence parameter, which was varied from $p_{0,G}$ to 130$\%$ of $p_{0,G}$. For comparison, the predicted fluxes from Poulin \textit{et al.}~\cite{PoulinVivianSalatiPhysRevD.99.023016} and Korsmeier \textit{et al.}~\cite{Korsmeier:2017xzj} and $\overline{\text{\textit{p}}}$ data from AMS-02~\cite{PhysRevLett.117.091103} are shown.

The $\overline{\text{\textit{p}}}$ flux predicted by this study exceeds the AMS-02 data by 20$\%$--30$\%$ in the low kinetic energy region (1--5\,GeV). This can be understood by looking at the comparison between the $\overline{\text{\textit{p}}}$ production cross sections used in this work (from {\tt EPOS-LHC}) and the di Mauro parametrization used by Poulin \textit{et al.} (Figure~\ref{fig:pbar_Xsec}). The differences in the low-energy region (below a few hundred GeVs in the laboratory frame) are especially important. Figure~\ref{fig:pbar_Xsec} (bottom left) shows that the $\overline{\text{\textit{p}}}$ production cross section ratios of {\tt EPOS-LHC} to the di Mauro parametrization reaches up to 1.2 for these low-energy collisions. Since low-energy collisions are the dominant source of antinuclei production in cosmic-ray interactions, the overproduction in {\tt EPOS-LHC} in this region is the major reason behind the excess $\overline{\text{\textit{p}}}$ flux predicted by this study.

The predicted secondary $\overline{\text{\textit{d}}}$ flux is very close to the predicted flux from~\cite{PoulinVivianSalatiPhysRevD.99.023016}. The predicted secondary ${}^3\overline{\text{He}}$ flux is consistently lower than the corresponding fluxes from both~\cite{PoulinVivianSalatiPhysRevD.99.023016} and~\cite{Korsmeier:2017xzj} by almost an order of magnitude, especially in the low kinetic energy region between 1 and 10\,GeV. As discussed in Sec.~\ref{s3}, the predicted ${}^4\overline{\text{He}}$ flux is shown only from 4--20\,GeV, and it agrees with~\cite{PoulinVivianSalatiPhysRevD.99.023016} within the uncertainties. 

It is important to note that the $p_0$ parametrization of Gomez \textit{et al.}~\cite{Gomez-Coral:2018yuk} for $\overline{\text{\textit{d}}}$ production, which has been extended and improved in this work for $\overline{\text{He}}$ production, already absorbs any differences between the $\overline{\text{\textit{p}}}$ production in {\tt EPOS-LHC} and experimental data. This is a consequence of the direct fit to $\overline{\text{\textit{d}}}$ and $\overline{\text{He}}$ data when extracting $p_0$. Hence, even if the excess in the predicted $\overline{\text{\textit{p}}}$ flux (discussed earlier) is corrected to match the AMS-02 data, the predicted fluxes of the heavier antinuclei shown in Figure~\ref{fig:sourceTerm} (right) will not be affected.

The differences in the antinuclei fluxes between this study and Ref.~\cite{PoulinVivianSalatiPhysRevD.99.023016} can be traced to the differences in the source terms in Figure~\ref{fig:sourceTerm} (left). The $\overline{\text{\textit{d}}}$ and ${}^3\overline{\text{He}}$ source terms are both smaller than the source terms in Ref.~\cite{PoulinVivianSalatiPhysRevD.99.023016}. This reduction observed at lower energies is a consequence of the energy-dependent $p_0$. The ${}^4\overline{\text{He}}$ source term is about the same within the uncertainties. However, due to the lack of experimental data for ${}^4\overline{\text{He}}$ production, the validity of the multiparticle coalescence model could not be evaluated for this regime.

\section{Conclusions}\label{s5}
To simulate the interaction of cosmic rays with the ISM, a multiparticle coalescence model was developed to produce light antinuclei in $p$-A collision simulations. A large-scale simulation of proton-proton collisions was carried out using this coalescence model, and the production cross sections of $\overline{\text{\textit{p}}}$, $\overline{\text{\textit{d}}}$, ${}^3\overline{\text{He}}$, and even ${}^4\overline{\text{He}}$ were estimated. These cross sections were validated at high energies of $\sqrt{s}$ = 7 and 13\,TeV by comparison with the latest data from ALICE, and also with $p$-Be and $p$-Al collisions at $\sqrt{s}$ = 19.4\,GeV. The lack of high-precision proton-proton experimental data at lower energies remains a crucial gap and affects the CR background predictions. 

The local source terms of these antinuclei were calculated, and a propagation model was used to predict the top-of-the-atmosphere secondary fluxes. These fluxes were compared to previous studies, which use a different methodology of scaling the $\overline{\text{\textit{p}}}$ cross section parametrizations to estimate the light antinuclei production cross sections.

The coalescence method developed here predicts about an order-of-magnitude lower antideuteron and antihelium fluxes than the numerical scaling models. In light of the AMS-02 antihelium candidate events, this study reinforces the prediction of extremely low antiparticle background for low-energy cosmic rays.

\section{Acknowledgments}\label{s6}
A. S., A. D., D. G., and P. v. D. would like to thank the National Science Foundation (Grant No. 1551980). C. K. thanks the DAAD RISE fellowship. 
This research was done using resources provided by the Open Science Grid~\cite{osg07, osg09}, which is supported by the National Science Foundation Grant No. 1148698, and the U.S. Department of Energy's Office of Science.
The technical support and advanced computing resources from the University of Hawaii Information Technology Services--Cyberinfrastructure are gratefully acknowledged.
The computing resources from the CERN batch system are also gratefully acknowledged.
We are grateful to Vivian Poulin and Pierre Salati for discussions and help with the galactic propagation code.

\clearpage
\pagebreak 

\bibliographystyle{unsrt}
\bibliography{bib_edit}

%
%

\end{document}